\documentclass[twocolumn,showpacs,preprintnumbers,amsmath,amssymb,superscriptaddress,footinbib]{revtex4}

\usepackage{graphicx}
\usepackage{dcolumn}
\usepackage{bm}

\begin{document}

\preprint{APS/123-QED}

\title{Excitation of a bosonic mode by electron tunneling into a cuprate superconductor} 

\author{P. Das}
  \email{Pintu.Das@cpfs.mpg.de}
\affiliation{Institute of Experimental Physics, University of
Saarland, D-66041 Saarbruecken, Germany} \affiliation{Max Planck
Institute of Chemical Physics of Solids, Noethnitzer Str. 40, 01187
Dresden, Germany.}
\author{M. R. Koblischka}
\affiliation{Institute of Experimental Physics, University of
Saarland, D-66041 Saarbruecken, Germany}
\author{H. Rosner}
\affiliation{Max Planck Institute of Chemical Physics of Solids,
Noethnitzer Str. 40, 01187 Dresden, Germany.}
\author{Th. Wolf}
\affiliation{Forschungszentrum Karlsruhe GmbH, Institute of Solid
State Physics, D-76021 Karlsruhe, Germany}
\author{U. Hartmann}
\affiliation{Institute of Experimental Physics, University of
Saarland, D-66041 Saarbruecken, Germany}
\date{\today}
\begin{abstract}
We performed scanning tunneling spectroscopic experiments on
hole-doped NdBa$_2$Cu$_3$O$_{7-\delta}$. The d$I$/d$V$ curves
obtained at 4.2 K are asymmetric with clear peak-dip and hump
structures. Energy derivatives of these curves show peaks at
energies beyond the dip features. Highly precise full potential
bandstructure calculations confirm a featureless electronic density
of states in that energy region. Our results indicate that tunneling
electrons couple to a collective mode in the CuO$_2$ plane.
\end{abstract}


\maketitle

The identification of the microscopic mechanism for pair formation
in the unconventional superconductors is still a challenge. In
conventional superconductors, apart from the isotope effect,
tunneling experiments provided the most direct evidence of
electron-phonon interaction mediating the Cooper-pair
formation~\cite{Rowell1962, McMillan1965}. In case of high-$T_c$
superconductors (HTSC), inelastic neutron scattering (INS)
experiments and high resolution angle-resolved photo emission
(ARPES) experiments were useful to find collective modes in the
range of 30-70 meV~\cite{Fong1999, Dai2001, Borisenko2006, Kim2003,
Lanzara2001, Cuk2004, Meevasana2006}. Collective spin excitations
have been suggested as the bosonic mode which interacts with
electrons to produce a resonance peak at a wavevector ($\pi,\pi$) in
INS spectra and a kink along the ($\pi$,0) direction of the
Brillouin zone in ARPES spectra~\cite{Fong1999, Dai2001, Sato2006,
Borisenko2006, Kim2003}. On the other hand, several ARPES results
also show evidence of phonons as the collective modes
\cite{Lanzara2001, Cuk2004, Meevasana2006}. Thus the assignment of
the boson-mediating Cooper pair formation in HTSC is far from being
settled.

Due to the microscopic electronic inhomogeneity in these
superconductors~\cite{Pan2001}, tunneling experiments using scanning
tunneling microscopy (STM) are expected to play an important role in
resolving such issues being related to pairing. While tunneling,
electrons can excite a collective mode of a certain energy if the
energy of the electrons is equal to that of the collective mode. Due
to the inelastic interaction of the electrons with the mode, a new
scattering channel is induced which results in a step-like feature
in the d$I$/d$V$ curves~\cite{Hansma1977}. Ideally, these step-like
features can be observed more easily in the form of a delta function
in the energy derivative of the d$I$/d$V$ curves. In experiments,
the peak heights and widths are dependent on the energy resolution,
temperature and the modulation voltage~\cite{Hansma1977}. For the
superconducting state of $d$-wave superconductors, Balatsky
\textit{et al.} predicted that features due to inelastic scattering
of tunneling electrons would be observed in the d$^2I$/dV$^2$ curves
as satellite peaks at $\Delta + \Omega$ in addition to a very weak
peak at $\Omega$, where $\Delta$ is the energy gap and $\Omega$ is
the excitation energy~\cite{Balatsky2003}. Indeed, Lee \textit{et
al.} have observed peaks in the second derivatives, which carry the
signature of a bosonic mode in the hole-doped cuprate
Bi$_2$Sr$_2$CaCu$_2$O$_{8-\delta}$ (Bi-2212)~\cite{Lee2006}. The
tunneling spectra of the cuprates like YBa$_2$Cu$_3$O$_{7-\delta}$
(Y-123), which contain an additional Cu-O layer in the unit cell in
the form of quasi-one-dimensional chains, often exhibited features
which were not observed for Bi-2212~\cite{Maggio1995, Ngai2007}.
These additional features in the spectra impeded the general
understanding of the electronic structure of the superconducting
state of cuprates and also masked the relevant features due to
bosonic mode from the tunneling spectra of these cuprates. Very
recently, Nietsemski \textit{et al.} observed a bosonic mode from
the tunneling spectra on Pr$_{0.88}$LaCe$_{0.12}$CuO$_4$ (PLCCO),
which is an electron-doped superconductor~\cite{Niestemski2007}. In
this work, we present evidence of a bosonic mode, which appears to
be originated at the CuO$_2$ plane, from scanning tunneling
spectroscopy (STS) data obtained on a twinned single crystal of
NdBa$_2$Cu$_3$O$_{7-\delta}$ (Nd-123). Nd-123 belongs to the Y-123
family. They have isostructural unit cells. The ionic size of
Nd$^{3+}$ ions, which are paramagnetic at 4.2
K~\cite{Boothroyd1999}, is bigger than that of Y$^{3+}$.

The single crystal used for the measurements was grown from a
BaO/CuO flux. The details of the growth process are according to
those mentioned in Ref.~\cite{Boothroyd1999}. After the growth, the
crystal was properly oxidized to achieve a $T_c$ onset of 93.5 K.
The transition width of $\Delta T_c \sim $ 3.5 K was determined from
ac-susceptibility measurements. For the STS experiments, the
as-grown crystal surface was cleaned with absolute ethanol and dried
by pure helium gas. No special surface preparation was performed
\footnote{Cleaved surfaces of Y-123 mostly expose the CuO chain
layer~\cite{Derro2002, Edwards1995}. In that case, a direct
tunneling to the CuO$_2$ plane layer, where superconductivity
originates, is usually not possible. On the other hand, STS
experiments on the as-grown surfaces sometimes reported to have
revealed features which are indicative of the CuO$_2$ plane-derived
spectra, see for example Ref.~\cite{Wu1999, Cren2000}. More
interestingly, Magio-Aprile \textit{et al.} have imaged vortices on
the as-grown single crystal surface of Y-123, see
Ref.~\cite{Maggio1995}.}. The STS measurements were carried out at
4.2 K in He gas atmosphere using a home-made STM. A mechanically cut
Pt-Ir tip was used for the measurements. The tunneling parameters
for the measurements were set as -0.1 V and 0.2 nA giving a
tunneling resistance of 500 M$\Omega$. The bias voltage was applied
to the sample so that a negative (positive) voltage refers to a
filled (empty) sample energy state. Spectroscopic data were obtained
as [d$I$/d$V$](V) curves using the standard lock-in technique where
a modulation voltage of 2 mV (rms) was used.
\begin{figure}
\begin{center}
\includegraphics[width=0.4\textwidth]{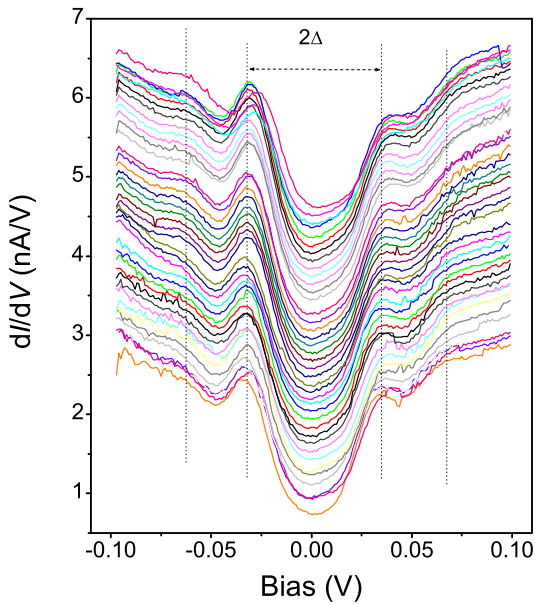}
\caption{(Color online) d$I$/d$V$ curves (shifted along y axis for
clarity) obtained at 4.2 K along a 20 nm line. The peaks and humps
are indicated by dotted vertical lines. \label{FigdIdVall}}
\end{center}
\end{figure}
Data were taken at many different locations on flat areas of the
sample surface. The spectra show that the surface is electronically
inhomogeneous. However, at some locations of the sample, we obtained
curves which are nearly homogeneous in the length scale of 20-25 nm.
Here we discuss only those spectra which contain coherence peaks
(P), dips (D) and humps (H). These features are hallmarks of the
superconducting state in cuprates as was independently observed in
STS and ARPES experiments on Bi-2212 \cite{DeWilde1998, Pan2000,
Campuzano1999, Kordyuk2002}. In Fig. \ref{FigdIdVall}, a set of
d$I$/d$V$ spectra obtained along a line cut of 20 nm is shown. These
spectra were observed while taking a d$I$/d$V$ curve at every 0.5 nm
of a 32 $\times$ 32 nm$^2$ area. It is remarkable that the PDH
features are observed on the as-grown sample surface. On the
as-prepared surface of NdBCO, Ting \textit{et al.} found the CuO
chain layer as the surface terminating layer~\cite{Ting1998}. The
spectra shown in Fig. 1 do not have any similarity to the ones
observed on the well characterized CuO chain layer~\cite{Derro2002}.
It is possible that the topmost CuO chain layer in the present case
is an insulating layer due to surface degradation. In that case, at
the small bias voltage range of $\pm$100 mV, the tunneling would
take place between the metal tip and the CuO$_2$ plane layer, which
is situated within $\sim$ 4.2 {\AA} of the CuO chain layer, thus
revealing the clear PDH features. Apart from the PDH features, the
asymmetry of the coherence peaks is evident from the data. The peak
height is found to be always larger at the filled sample states
compared to the empty ones. Similar asymmetric curves were also
observed for Bi-2212~\cite{DeWilde1998, Sugita2000,Pan2001} and
Nd-123~\cite{Wu1999} earlier. The asymmetry is most likely related
to the probability of electron extraction and injection from/to the
material~\cite{Anderson2006}. Beyond the peaks, there is an
asymmetric V-shaped background on top of which humps are evident
both at empty and filled states. The humps are observed at $\sim$
2$\Delta$ and there are dips at $\sim$ 1.4$\Delta$. The respective
features are broader for the empty states than for the filled ones.
This becomes clear from the representative curve shown in Fig.
\ref{FigdIdV1}. The average energy gap measured from peak to peak is
70 meV which leads to 2$\Delta/k_BT_c$ = 8.68.

\begin{figure}[b]
\begin{center}
\includegraphics[width=0.5\textwidth]{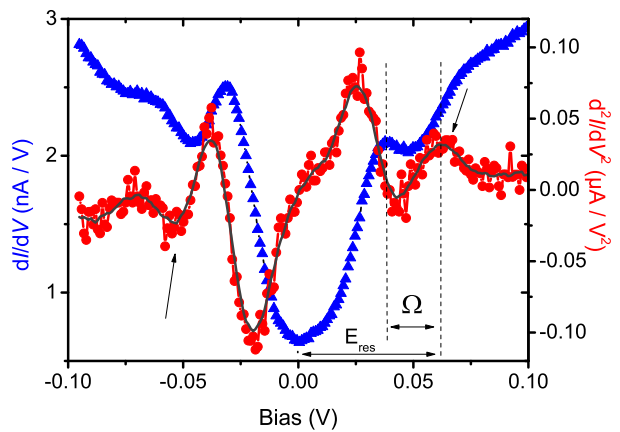}
\caption{(Color online) Typical d$I$/d$V$ curve (triangles)
containing peak-dip-hump structures together with  the energy
derivative (circles). The smooth dark line is a guide to the eyes.
The arrows show the inelastic features at both empty and filled
states. The boson mode energy ($\Omega$) at empty state is shown.
\label{FigdIdV1}}
\end{center}
\end{figure}

Recently, Ngai \textit{et al.} reported subgap features in addition
to the coherence peaks in the d$I$/d$V$ curves obtained on
Y$_{1-x}$Ca$_x$Ba$_2$Cu$_3$O$_{7-\delta}$ thin films
~\cite{Ngai2007}. Similar features were also reported earlier on
Y-123 single crystals~\cite{Maggio1995}. On the other hand, subgap
features were not observed in the ARPES data on cleaved single
crystal and STS data on thin film of Y-123~\cite{Lu2001, Cren2000}.
However, the PDH features are very clearly present in those data.
The subgap features were also not observed by Wu \textit{et al.} on
Nd-123~\cite{Wu1999}. In cuprate systems with an additional
copper-oxygen chain layer, the subgap features were suggested to be
due to multiband superconductivity~\cite{Ngai2007}. In our data, we
do not find any indication of such an influence of the metallic
chain layer on the electronic structure representing superconducting
state (see Fig. \ref{FigdIdVall}). Interestingly, Derro \textit{et
al.} observed strong subgap resonances in the energy scale of $\sim$
5 meV while tunneling to the chain layer of Y-123 single
crystals~\cite{Derro2002}. These subgap features were suggested to
be possibly due to the oxygen vacancies in the chains. It is most
likely that the metallic CuO chain layers acquire superconductivity
due to the proximity effect~\cite{Derro2002, Morr2001}. Within this
scenario, the spectra with subgap features possibly reflect the LDOS
of CuO chains and those with clear PDH features reflect the LDOS of
CuO$_2$ plane layers as these features are also observed on
chain-less Bi-2212. If the superconductivity in 123 systems would be
due to multiband effects, the signature of this should also be seen
in the electronic spectra of Nd-123. Although from the present data
it is not clear if the subgap states would be observed in the LDOS
of CuO$_2$ plane that exists in the bulk, however, it gives another
indication that the PDH features are the generic features in the
CuO$_2$ plane-derived DOS of cuprate
superconductors. 

In order to gain further information from the spectra, we focus on
the part of the curves beyond the coherence peaks. By numerically
differentiating the curves we observe a peak in the empty state at
an energy $E_{res}$ as shown in Fig. \ref{FigdIdV1}. The above
mentioned peaks in the d$^2I$/d$V^2$ curves correspond to very weak
step-like features in the d$I$/d$V$ curves. The observed fine
structure is consistent with the predicted inelastic electron tunnel
spectroscopic (IETS) features in $d$-wave
cuprates~\cite{Balatsky2003, Zhu2006}. It also has considerable
similarity to the IETS features observed for hole-doped
Bi-2212~\cite{Lee2006} and electron-doped PLCCO
superconductors~\cite{Niestemski2007}. In order to check if the fine
structure in the STS spectra is due to the specific bandstructure of
the superconductor, full potential density functional calculations
were performed. We applied the full potential local-orbital minimal
basis scheme FPLO~\cite{Koepernik1999} (version 5.00-19) within the
local spin density approximation (LSDA), where the exchange
correlation potential of Perdew-Wang~\cite{Perdew1992} was chosen
for the calculations. The strong Coulomb repulsion in the localized
Nd $4f$ shell has been modeled in a mean-field way within the
LSDA+$U$ approximation. A typical value of $U_{4f}$ = 8 eV has been
applied throughout the calculations. Our results are basically
independent of the choice of $U$ within physically reasonable
limits. As basis set, Nd $4f5s5p/6s6p5d$, Ba $5s5p/6s6p5d$, Cu
$3s3p/4s4p3d+4d$ and O $2s2p3d+3s$ states (notation:
semicore/valence+polarization states) have been chosen. The extent
of the valence states is optimized with respect to the total
energy~\cite{Eschrig2003}. The inclusion of the semicore states is
necessary to account for their nonnegligible overlap. The
polarization states provide a more complete basis set to insure
highly accurate bandstructure and DOS results. A very fine $k$-mesh
of 16200 points in the Brillouin zone (2560 in the irreducible
wedge) was used to resolve the singularities in the DOS near the
Fermi level accurately. Convergency with respect to basis set and
$k$-mesh was carefully checked.
\begin{figure}
\begin{center}
\includegraphics[width=0.5\textwidth]{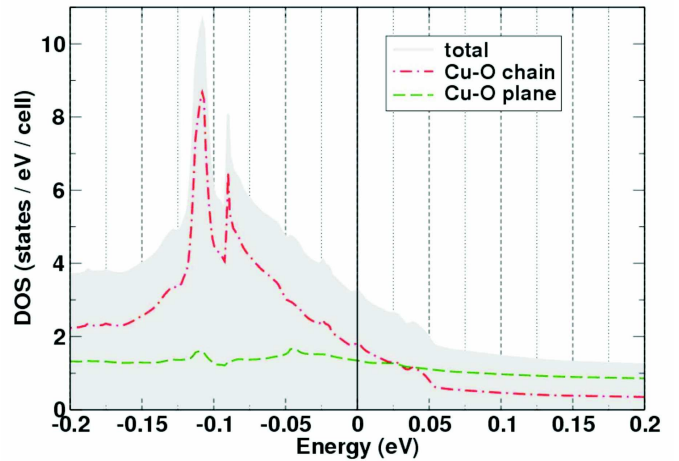}
\caption{(Color online) Calculated total electronic density of
states (shaded gray) in the vicinity of the Fermi level (zero
energy) for NdBCO. The sharp features at about 85 meV and 110 meV
binding energy can be assigned to van-Hove singularities originating
from the Cu-O chains ([red] dot-dashed line). The contribution from
the plane ([green] dashed line) is rather flat in the shown energy
window. \label{FigDOS}}
\end{center}
\end{figure}
The structural data of Ref.~\cite{Wuernisha2006} were used.

We find a total valence band width of about 9 eV, typical for the
quasi-two-dimensional cuprates. Three bands cross the Fermi level,
two of them originating from the CuO$_2$ layer and one originating
from the CuO$_3$ chains along the $y$ direction. In Fig.
\ref{FigDOS}, we show the calculated total DOS of Nd-123 in the
vicinity of the Fermi level. In this energy region, only the states
originating from the CuO$_2$ planes and the CuO$_3$ chains exhibit
significant contributions. Whereas the CuO$_2$-planes-related DOS is
almost constant close to the Fermi level, two distinct van Hove
singularities related to the CuO$_3$ chain show up at about 85 and
110 meV. Otherwise, no significant features in the DOS near the
Fermi level are obtained. The slight underdoping of the present
sample can be modeled by a small rigid shift of the Fermi level
towards negative energies. For an oxygen deficiency $\delta$ = 0.01,
this shift is estimated to be $\sim$ -18 meV. There is no sharp
structure present in the DOS of the CuO$_2$ plane at energies close
to the region where peaks are observed in the experimental
d$^2I$/d$V^2$ spectra. Thus, these features cannot be related to the
DOS. The calculations also show that the bilayer splitting is not
responsible for the PDH features.

\begin{figure}
\begin{center}
\includegraphics[width=0.5\textwidth]{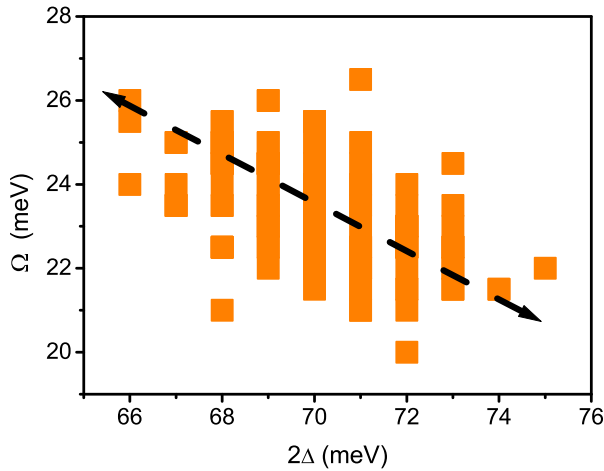}
\caption{(Color online) Mode energy ($\Omega(r)$) vs energy gap
($2\Delta(r)$) obtained from the d$I$/d$V$ curves like in Fig. 2.
The dashed line emphasizes the tentative correlation. The
uncertainties in the evaluation of $\Delta$ and $\Omega$ are $\pm
(0.5 \cdots 1)$ meV and $\pm (1 \cdots 1.5)$ meV,
respectively.\label{FigCorr}}
\end{center}
\end{figure}

In accordance with Eliashberg's classical theory for strong coupling
superconductors, where the electron-boson interaction was proposed
to be observed at an energy of $E = \Delta +
\Omega$~\cite{Eliashberg1960}, the $\Omega$ values were determined
for both empty and filled state. The empty and filled state $\Omega$
are experimentally determined by $\Omega(r)$ = $E_{res}(r) -
\Delta(r)$ from the entire set of curves obtained at different
locations (r) on the sample. The mean values of $\Omega(r)$ are
found to be 22.9 $\pm$ 1.8 meV and 23.7 $\pm$ 1.6 meV for empty and
filled states, respectively. Thus, the peaks at $E_{res}$ in the
d$^2I$/d$V^2$ curves are found to be almost symmetric. This strongly
supports the assumption that the observed features are signatures of
inelastic tunneling. They result from an additional tunneling
channel introduced due to the excitation of a collective mode at
$\sim$ 23 meV. The mode energy is surprisingly low compared with the
ones observed for Y-123 for similar dopings~\cite{Dai2001}. A
priori, it is not clear if the mode is excited in the tunneling
barrier or in the CuO$_2$ plane of the sample. Thus, it is
instructive to plot $\Omega(r)$ against 2$\Delta(r)$. As shown in
Fig. \ref{FigCorr}, $\Omega$ and $\Delta$ are intimately related. A
similar, but comparatively stronger, correlation was also found for
Bi-2212 and PLCCO~\cite{Lee2006, Niestemski2007}. Moreover, this
mode energy for Nd-123 was observed for only those spectra which
exhibit clear PDH features. Thus, the collective mode appears to be
intrinsic to the superconducting CuO$_2$ plane.
To determine the nature of the observed collective mode, we compare
the mode energy with the known phonon energies of Nd-123. The former
is rather low compared to the in-phase or out-of-phase O(2)-O(3)
vibrations which have energies of more than 35 meV as observed from
Raman spectra~\cite{Heyen1991, Misochko1997, Liu1988}. In case of
Bi-2212, Pilgram \textit{et al.} suggested that the peaks in the
second derivative are due to the excitation of an apical oxygen
vibration mode~\cite{Pilgram2006}. However, the corresponding energy
for Nd-123 as detected from Raman spectra is $\sim$ 68
meV~\cite{Heyen1991}. Also, this is not intrinsic to the CuO$_2$
plane. The only low-energy modes for Nd-123 are due to the metallic
Ba (14 meV) and Cu (19 meV) ion vibration modes. The latter one
seems to be the only possible mode that could be excited during
tunneling in the present case. However, it is not clear why in that
case the same mode energy is not observed in case of Bi-2212 or
PLCCO~\cite{Lee2006, Niestemski2007}. Thus, although the mode does
not appear to be a phonon, we can not definitely rule out the
possibility of a low energy excitation of metal ions.

For an underdoped Y-123 sample, incommensurate spin fluctuations
were observed at 24 meV in INS experiments~\cite{Mook1998}.
Niestemski \textit{et al.} have suggested the 10.5 meV mode as
observed for PLCCO to be a magnetic resonance
mode~\cite{Niestemski2007}. At this point, it remains unclear what
the origin of the observed mode for this rare-earth-based cuprate
is. Since the features observed in the spectra and the correlation
between $\Delta$ and $\Omega$ are qualitatively similar to those
observed for Bi-2212, the identification of the nature of this low
energy mode in the rare-earth-based cuprate would be important.

In summary, our STS data on as-grown slightly underdoped Nd-123
single crystal do not show any evidence of multiband
superconductivity. However, it would be important to probe the
CuO$_2$ plane layer in the bulk in order to completely rule out the
role of CuO chain band towards superconductivity in these systems.
The peak-dip-hump features in the spectra indicate that these are
most likely the generic features of the CuO$_2$ plane-derived LDOS
in cuprates. Furthermore, a bosonic mode has been detected with a
characteristic mode energy of about 23 meV. The mode energy is
rather low compared to the other bosonic modes observed for the
cuprates. The observed mode energy has certain correlation with the
energy gap and thus appears to be intrinsic to CuO$_2$ plane.
However, the correlation is weak compared to the ones reported for
other cuprates. Thus, the origin of the low energy mode due to a
tunneling pathway, which is extrinsic to the CuO$_2$ plane layer,
can not be completely ruled out.

We acknowledge the useful discussions with J.-X Zhu, A. V. Balatsky,
S. V. Borisenko, H. Ghosh and D. Scalapino. One of the authors (HR)
acknowledges financial support from the Emmy Noether program of the
Deutsche Forschungsgemeinschaft.

\bibliography{PRL.bbl}

\end{document}